\documentclass[preprint2]{aastex}

\newcommand{\kt}{\ensuremath{k_{\rm{B}}T}}
\newcommand{\lx}{\ensuremath{L_{\rm{X}}}}
\newcommand{\lbol}{\ensuremath{L_{\rm{bol}}}}

\newcommand{\fx}{\ensuremath{F_{\rm{X}}}}
\newcommand{\fk}{\ensuremath{F_{K_{\rm{S}}}}}
\newcommand{\lk}{\ensuremath{L_{K_{\rm{S}}}}}
\newcommand{\fxc}{\ensuremath{F_{\rm{X}}^{\rm{cor}}}}
\newcommand{\nh}{\ensuremath{N_{\rm H}}}
\newcommand{\ji}{\textit{J}}
\newcommand{\hi}{\textit{H}}
\newcommand{\ki}{\textit{K}}
\newcommand{\ksi}{\textit{K\ensuremath{_S}}}

\slugcomment{}
\shorttitle{X-ray \& NIR Studies of L1448}
\shortauthors{Tsujimoto, Kobayashi, \& Tsuboi}

\begin{document}
\title{X-ray and Near-infrared Studies of a Star-forming Cloud; L1448}
\author{M.~Tsujimoto\altaffilmark{1,2}, N.~Kobayashi\altaffilmark{3}, \& Y.~Tsuboi\altaffilmark{4}}
\altaffiltext{1}{Visiting Astronomer, Kitt Peak National Observatory, National Optical
Astronomy Observatory, which is operated by the Association of Universities for Research
in Astronomy, Inc. (AURA) under cooperative agreement with the National Science
Foundation.}
\altaffiltext{2}{Department of Astronomy \& Astrophysics, The Pennsylvania State
University, 525 Davey Laboratory, University Park, PA 16802.}
\altaffiltext{3}{Institute of Astronomy, The University of Tokyo, 2-21-1 Osawa, Mitaka,
Tokyo 181-0015, Japan.}
\altaffiltext{4}{Department of Physics, Chuo University, Kasuga 1-13-27, Bunkyo-ku,
Tokyo 112-8551, Japan.}

\begin{abstract}
 We present the results of X-ray and near-infrared (NIR) observations of L1448, a
 star-forming region in the Perseus cloud complex using the \textit{Chandra} X-ray
 Observatory and the 4 m telescope at the Kitt Peak National Observatory. We detect 72
 X-ray sources in a $\sim$17\arcmin$\times$17\arcmin\ region with a $\sim$68~ks ACIS
 exposure, for which we conduct follow-up NIR imaging observations in a concentric
 $\sim$11\arcmin$\times$11\arcmin\ region with FLAMINGOS down to $m_{K_{S}}$
 $\sim$~17~mag. Twelve X-ray sources have NIR or optical counterparts. By plotting X-ray
 mean energy versus NIR to X-ray flux ratio, the X-ray sources are clearly separated
 into two groups. The X-ray spectral and temporal features as well as NIR magnitudes and
 colors indicate that one group mainly consists of young stellar objects (YSOs) in the
 cloud and the other of background extragalactic sources. Ten X-ray-emitting YSO
 candidates are thus newly identified, which are low-mass or brown dwarf mass sources
 from their NIR magnitudes. In addition, a possible X-ray signal is found from a
 mid-infrared protostar L1448 IRS~3(A). The lack of detection of this source in our deep
 NIR images indicates that this source has a very steep spectral slope of $\gtrsim$3.2
 in 2--10~\micron.
\end{abstract}

 
\section{INTRODUCTION}
L1448 is a dark cloud \citep{lynds62} located at the western edge of the Perseus
molecular cloud complex at a distance of $\sim$300~pc. It has a size of
$\sim$15\arcmin$\times$7\arcmin\ in a C$^{18}$O map and a mass of $\sim$100~$M_{\odot}$
\citep{bachiller86}. The region has been intensively studied in wavelengths longer than
far-infrared focusing mainly on a giant molecular outflow, and only few studies in the
near-infrared (NIR) broad-band \citep{bally93} and none in the X-ray band have been
presented, both of which are currently strong measures to identify stellar constituents
of star-forming clouds. The infrared excess due to emission from circumstellar material
and the X-ray brightness are two major observational characteristics of young stellar
objects (YSOs), which can be respectively revealed through NIR broad-band photometry in
the \ji, \hi, and \ksi\ bands and X-ray imaging observations. The purpose of this paper
is to report the results of the first X-ray imaging-spectroscopy study of L1448 aided by
follow-up NIR broad-band imaging observations. Previous X-ray studies in the Perseus
cloud complex can be found in \citet{preibisch96,preibisch01,preibisch02,preibisch04b}
for IC\,348, \citet{getman02,preibisch03} for NGC\,1333, and \citet{yamauchi01} for
Barnard 1 and IRAS\,03282$+$3035.

\smallskip
The most magnificent feature in L1448 is a parsec-scale bipolar outflow
emanating from the core of the cloud, which is one of the
highest-velocity, best-collimated, and youngest outflows known \citep{davis95}. The
outflow was discovered by \citet{bachiller90} in a molecular emission line image. The
exciting source of the outflow was promptly detected in the millimeter and centimeter
continua \citep{bachiller91,curiel90}, which is referred as L1448 mm or L1448
C(enter). \citet{barsony98} obtained a spectral energy distribution (SED) in
far-infrared to millimeter wavelengths and established the class~0 status of this
source.

The \textit{Infrared Astronomical Satellite} (IRAS) detected three sources (IRAS
03220$+$3035, 03222$+$3034, and 03225$+$3034) in this region, which are called L1448
IRS~1, 2, and 3. IRS~1 has a visual counterpart with $m_{V}$ $=$ 19.5~mag
\citep{cohen79}. From its strong H$\alpha$ emission and excess emission in the
\textit{IRAS} band \citep{cohen79,degrijp87}, it is probably a classical T Tauri star or
a class~I protostar in the cloud. IRS~2 is a class~0 source confirmed by
\citet{olinger99} based on an SED using \textit{IRAS} and SCUBA data. This source is
associated with various signatures of extreme youth, including CO and H$_{2}$ molecular
outflows \citep{olinger99,eisloeffel00}, Herbig-Haro objects \citep{bally97}, an
H$_{2}$O maser source \citep{anglada89}, and a centimeter continuum emission
\citep{anglada02}. IRS~3, also known as L1448 N(orth), is resolved into three
components; IRS~3(A), 3(B), and 3(C), also known respectively as L1448 N(A), N(B), and
NW. The former two constitute a binary separated by $\sim$ 2100~AU. Classification of
the individual binary components in millimeter to far-infrared observations had been
difficult due to crowding. However, recent mid-infrared (MIR) observations clearly
resolved IRS~3(A) and 3(B) with an \textit{N}-band detection only from 3(A). Based on
this, \citet{ciardi03} argued that 3(B) is a class 0 protostar whereas 3(A) might be at
the transition phase between class 0 and I stages. IRS~3(C), away from the binary by
$\sim$20\arcsec, is another class~0 source \citep{barsony98}.

\section{OBSERVATIONS}
The X-ray observation was carried out on 2004 March 15 using ACIS (Advanced CCD Imaging
Spectrometer; \citealt{garmire03}) on-board \textit{Chandra} \citep{weisskopf02}. Four
ACIS-I chips (I0, I1, I2, and I3) covered a $\sim$17\arcmin$\times$17\arcmin\ region
aimed at R.\,A.\,$=$\,3$^{\rm h}$25$^{\rm m}$36.4$^{\rm s}$ and
Decl.\,$=$\,30\arcdeg44\arcmin58\arcsec\ (J2000.0) for an exposure time of
67.9~ks. ACIS has sensitivity in the 0.5--9.0~keV band with a resolution of
$\sim$150~eV at 6~keV and a superb point spread function (PSF) radius of
$\sim$0.5\arcsec\ at the on-axis position. The data were taken with the very faint
telemetry mode and the timed exposure CCD operation with a frame time of 3.2~s.

\begin{figure}[hbtp]
 \figurenum{1}
\epsscale{1.0}
 \plotone{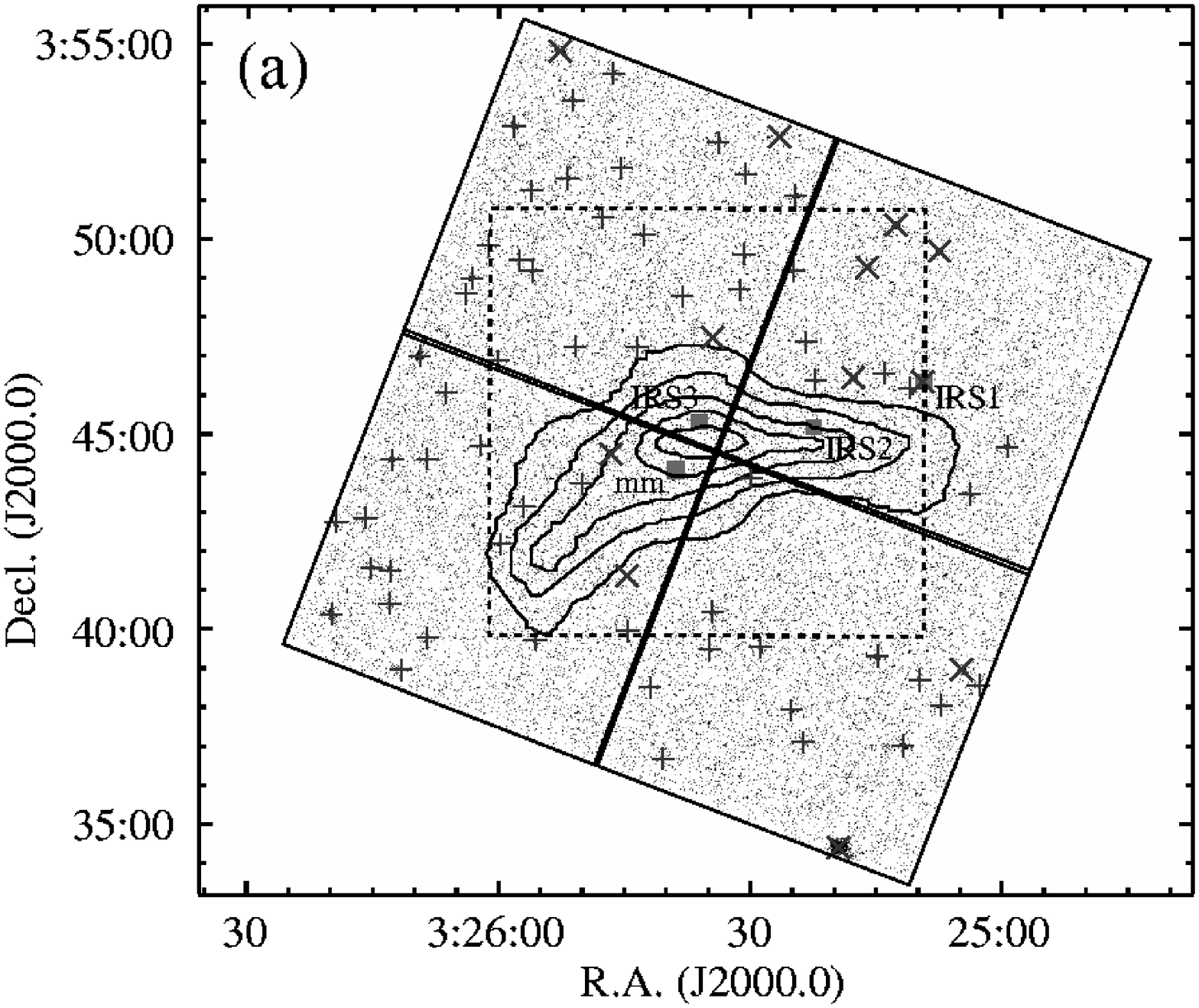}
 \plotone{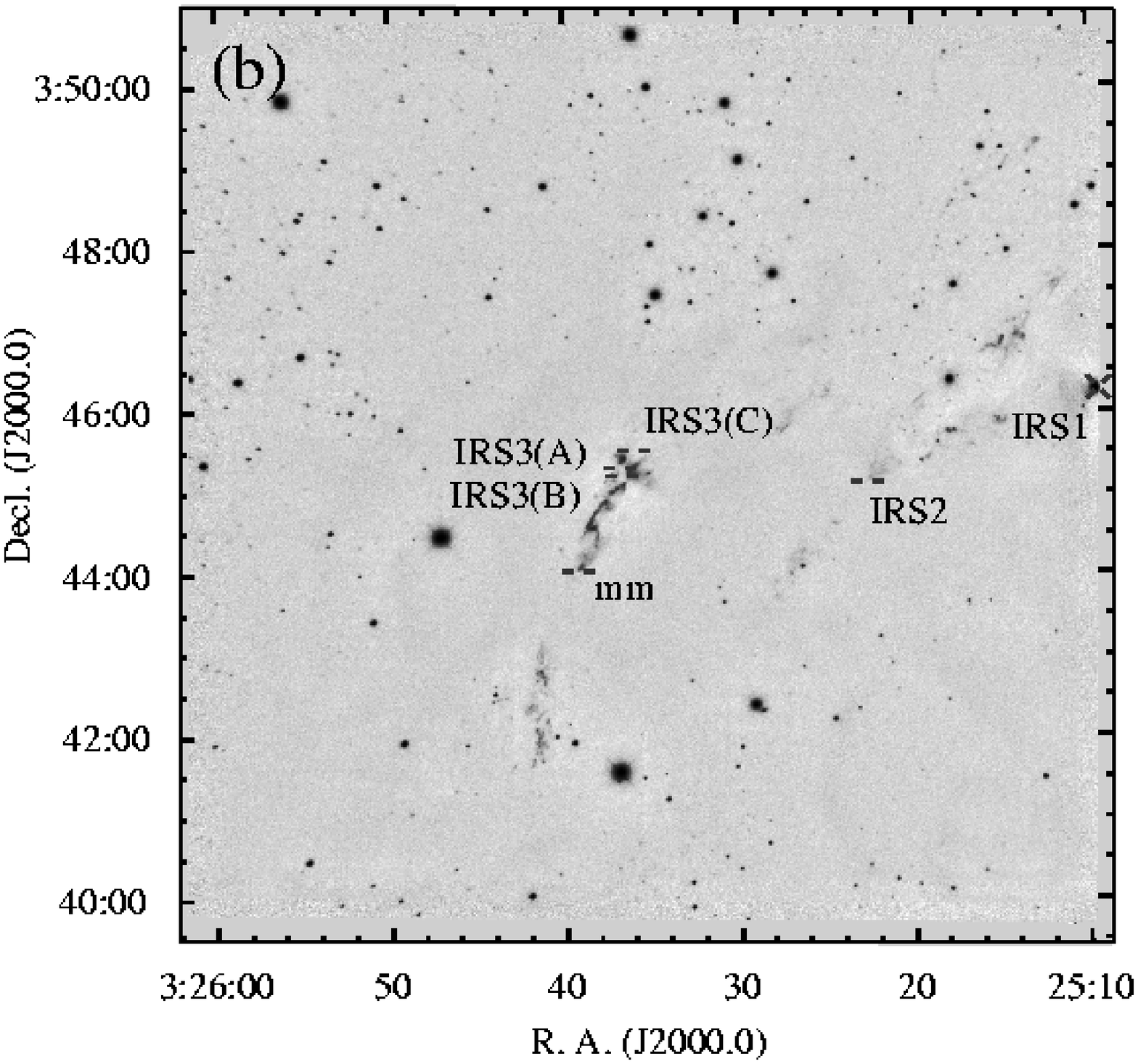}
 \caption{(\textit{a}) X-ray image. The 0.5--9.0~keV intensity is given by a logarithmic
 grey scale. The positions of X-ray sources with and without NIR counterparts are shown
 by crosses and pluses respectively, whereas the position of three IRAS sources
 (IRS~1--3) and L1448 mm are by the filled squares. The contours indicate the intensity
 of an NH$_{3}$ transition line \citep{anglada89}. The fields of view of four ACIS-I
 chips and FLAMINGOS are illustrated with the solid oblique squares and the dotted
 square. (\textit{b}) NIR image. The \ksi-band intensity is given by a logarithmic grey
 scale. The positions of embedded sources are bracketed with a pair of solid lines when
 undetected and by crosses when detected.}\label{fg:f1}
\end{figure}

The NIR observation was conducted on 2003 December 16 using FLAMINGOS \citep{elston03}
on the Cassegrain focus of the 4 m telescope at the Kitt Peak National
Observatory. FLAMINGOS uses a HgCdTe HAWAII-2 array which has a format of
2048$\times$2048 with a pixel scale of $\sim$0.316\arcsec~pixel$^{-1}$. With dithering
observations of an amplitude of 15\arcsec, we obtained \ji-, \hi-, and \ksi-band images
that cover a $\sim$11\arcmin$\times$11\arcmin\ area centered close to the ACIS aim point
at R.\,A.\,$=$\,3$^{\rm h}$25$^{\rm m}$35.1$^{\rm s}$ and
Decl.\,$=$\,30\arcdeg45\arcmin19\arcsec\ (J2000.0). The total exposure time was 13.5, 7,
and 17 minutes respectively in the \ji, \hi, and \ksi\ bands. The seeing was
$\sim$1\arcsec. Figures~\ref{fg:f1} (\textit{a}) and (\textit{b}) show the X-ray and NIR
intensity maps. The outflow from L1448 mm appears clearly in the \ksi\ band
(Fig.~\ref{fg:f1}\textit{b}) presumably due to H$_{2}$~$v$=1--0~S(1),
vibrational-rotational transition line emission by excited hydrogen molecules.

\section{DATA REDUCTION \& ANALYSIS}
\subsection{X-ray Data}
We reprocessed the Level 1 data distributed by the \textit{Chandra} X-ray Center to
obtain our X-ray event list. The latest calibration results (CALDB 2.28) were
incorporated and the background rejection algorithm specific to data taken with the very
faint mode was applied. Events were further cleaned by removing cosmic-ray afterglows
and applying filters based on the event grade, status, and good-time intervals.

We detected sources using a wavelet technique in CIAO\footnote{See
http://asc.harvard.edu/ciao/ for detail.} from three images of different energy bands;
soft (0.5--2.0~keV), hard (2.0--9.0~keV), and total (0.5--9.0~keV). For each detected
source, we used ACIS Extract\footnote{See
http://www.astro.psu.edu/xray/docs/TARA/ae\_users\_guide.html for detail.} for a
systematic event extraction, background subtraction, calculation of instrumental
responses, and construction and binning of spectra and light curves. Source events
were extracted from a 90\% encircled energy polygon of the 1.5~keV PSF, whereas
background events were locally extracted from an annular region around the source. The
positions of sources were fine-tuned by correlating the event distribution with the
PSF. As the X-ray source distribution is relatively sparse in the study field
(Fig.~\ref{fg:f1}\textit{a}), we do not suffer any overlaps of extraction regions. The
details of the procedure are described in \citet{getman05}.

For each source, we derived X-ray photometry information consisting of the source
position and its uncertainty, source count (0.5--8.0~keV), net count rate (NCR),
probability of no source (PNS), photometry significance (PS), and mean energy
(ME). The net count ($C_{\rm{net}}$) is calculated from the number of counts
($C_{\rm{src}}$ and $C_{\rm{bkg}}$) and the extraction area ($A_{\rm{src}}$ and
$A_{\rm{bkg}}$) of the source and background regions as $C_{\rm{net}} =
C_{\rm{src}}-C_{\rm{bkg}}\times(A_{\rm{src}}/A_{\rm{bkg}})$, which is divided by the
effective exposure time to derive NCR. PNS is an index for the detection significance,
with which we test the null hypothesis that all detected counts are explained by
background fluctuation;
\begin{equation}
 \mathrm{PNS} = 1 - \int_{0}^{C_{\rm{src}}-1}dN~P(C_{\rm{bkg}}\left(\frac{A_{\rm{src}}}{A_{\rm{bkg}}}\right),N),
\end{equation}
where $P(\lambda,N)$ represents a Poisson probability distribution function of a mean
$\lambda$ to have $N$ counts. PS is a metric for the reliability of photometry defined
as
\begin{equation}
 \mathrm{PS} = \frac{C_{\mathrm{net}}}{\sqrt{(\Delta C_{\mathrm{src}})^{2}+(\Delta C_{\mathrm{bkg}})^{2}\times(A_{\mathrm{src}}/A_{\rm{bkg}})^{2}}},
\end{equation}
where the approximation by \citet{gehrels86} is used to estimate $\Delta
C_{\rm{src}}=1+\sqrt{C_{\rm{src}}+0.75}$ and $\Delta
C_{\rm{bkg}}=1+\sqrt{C_{\rm{bkg}}+0.75}$. ME is a metric to represent the spectral
hardness of sources. The value is the average incident energy ($E$) weighted by
$C_{\rm{net}}$ of each energy bin and is defined as
\begin{equation}
 \mathrm{ME} = \frac{\int dE~E~C_{\rm{net}}(E)}{\int dE~C_{\rm{net}}(E)}.
\end{equation}
Table~\ref{tb:t1} shows the results of the ACIS photometry, consisting of 72 sources with
high significance both in the detection (PNS $\le$ 1$\times$10$^{-3}$) and the
photometry (PS $\ge$ 2).

\begin{deluxetable}{lccrrrrrrc}
 \tabletypesize{\scriptsize}
 \tablecaption{ACIS Photometry\label{tb:t1}}
 \tablecolumns{10}
 \tablewidth{0pt}
 \tablehead{
 \colhead{X-ray\tablenotemark{a}} &
 \colhead{R.A.\tablenotemark{b}} &
 \colhead{Decl.\tablenotemark{b}} &
 \colhead{Error} &
 \colhead{Counts} &
 \colhead{NCR\tablenotemark{c}} &
 \colhead{PNS\tablenotemark{c}} &
 \colhead{PS\tablenotemark{c}} &
 \colhead{ME\tablenotemark{c}} &
 \colhead{NIR\tablenotemark{d}} \\
 \colhead{ID} &
 \colhead{(hh:mm:ss.s)} &
 \colhead{(dd:mm:ss)} &
 \colhead{(\arcsec)} &
 \colhead{} &
 \colhead{(ks$^{-1}$)} &
 \colhead{} &
 \colhead{} &
 \colhead{(keV)} &
 \colhead{Counterpart} 
 }
 \startdata
1 & 3:24:59.3 & 30:44:41 & 0.7 & 25 & 0.3 & 0 & 3.2 & 3.6 & \nodata \\
2 & 3:25:02.7 & 30:38:34 & 1.1 & 25 & 0.6 & 9e$-$7 & 2.8 & 3.3 & \nodata \\
3 & 3:25:03.8 & 30:43:28 & 0.6 & 20 & 0.3 & 0 & 3.0 & 2.7 & \nodata \\
4$^{\dagger}$ & 3:25:04.8 & 30:38:58 & 0.4 & 116 & 2.4 & 0 & 9.0 & 1.4 & 03250474$+$3038578 \\
5 & 3:25:07.2 & 30:38:04 & 1.0 & 34 & 0.6 & 0 & 3.8 & 3.6 & \nodata \\
  \enddata
 \tablenotetext{a}{Variable sources are marked with a dagger.}
 \tablenotetext{b}{Equinox in J2000.0.}
 \tablenotetext{c}{See text for definitions of NCR, PNS, PS, and ME. PNS values are
 replaces with 0 when PNS $<$ 1e$-$9.}
 \tablenotetext{d}{2MASS counterparts are given in the format of hhmmssss$+$ddmmsss, while
 FLAMINGOS counterparts are given by their source numbers in Table~\ref{tb:t3}.}
 \tablecomments{The complete version of this table is in the electronic
  edition of the Journal.  The printed edition contains only a sample.}
\end{deluxetable}

For each source, flux variability was tested using the K-S test and the Bayesian block
segmentation technique \citep{scargle98}. We recognized flux to be variable if the number
of Bayesian blocks exceeds two and the K-S probability is less than 1\%. Four sources
(Nos. 4, 7, 34, and 69) have variable flux, which are marked with daggers in
Table~\ref{tb:t1}.

For 18 sources with more than 40 counts, we performed spectral analysis. The spectra
were fit with a one-temperature thin-thermal plasma model (the APEC model;
\citealt{smith01}) with the metallicity value fixed to 0.3 solar to determine the
best-fit absorption (\nh), plasma temperature (\kt), and X-ray luminosity (\lx)
values. They were also fit with a power-law model to derive the best-fit \nh, index
($\Gamma$), and absorption-corrected flux (\fxc) values. Table~\ref{tb:t2} shows the
results of 13 and 11 successful fits with the thermal and power-law models with a
reasonable value of \kt$<$~10~keV and $\Gamma <$~3.

\begin{deluxetable}{rcccccccc}
 \tabletypesize{\scriptsize}
 \tablecaption{ACIS Spectroscopy of Bright Sources\label{tb:t2}}
 \tablecolumns{8}
 \tablewidth{0pt}
 \tablehead{
 \colhead{} &
 \multicolumn{3}{c}{-------------------- thermal model --------------------} &
 \colhead{} &
 \multicolumn{3}{c}{------------------- power-law model -------------------} \\
 \colhead{X-ray} &
 \colhead{\nh\tablenotemark{a}} &
 \colhead{\kt\tablenotemark{a}} &
 \colhead{\lx\tablenotemark{a,b}} &
 \colhead{} &
 \colhead{\nh\tablenotemark{a}} &
 \colhead{$\Gamma$\tablenotemark{a}} &
 \colhead{\fxc\tablenotemark{a,b}} \\
 \colhead{ID} &
 \colhead{(10$^{22}$ cm$^{-2}$)} &
 \colhead{(keV)} &
 \colhead{(10$^{27}$ ergs~s$^{-1}$)} &
 \colhead{} &
 \colhead{(10$^{22}$ cm$^{-2}$)} &
 \colhead{} &
 \colhead{(10$^{-14}$ ergs~s$^{-1}$~cm$^{-2}$)} 
 }
 \startdata
4 & 0.0 (0.0--0.1) & 1.7 (1.5--1.9) & 1.8 (1.1--2.0) & & \nodata & \nodata & \nodata \\
6 & 0.4 (0.2--0.6) & 0.9 (0.7--1.1) & 1.3 (0.0--1.6) & & \nodata & \nodata & \nodata \\
10 & 1.3 (0.8--2.0) & 2.7 (1.4--5.9) & 3.6 (0.2--3.8) & & 1.6 (1.3--2.3) & 2.7 (2.3--3.3) & 5.7 (0.0--6.7) \\
11 & \nodata & \nodata & \nodata & & 0.3 (0.2--0.6) & 1.2 (0.9--1.5) & 4.1 (0.0--4.6) \\
15 & 0.0 (0.0--0.2) & 1.0 (0.9--1.2) & 1.0 (0.0--1.5) & & \nodata & \nodata & \nodata \\
16 & 0.3 (0.2--0.4) & 5.3 (4.3--7.1) & 24.9 (21.7--26.5) & & 0.4 (0.3--0.5) & 1.9 (1.8--2.0) & 26.1 (21.7--29.0) \\
20 & 0.3 (0.2--0.3) & 7.9 (5.5--11.2) & 10.2 (8.3--10.7) & & 0.3 (0.3--0.4) & 1.7 (1.6--1.8) & 10.1 (7.8--11.4) \\
23 & 0.3 (0.1--0.5) & 0.6 (0.5--0.8) & 0.9 (0.4--1.3) & & \nodata & \nodata & \nodata \\
42 & 0.0 (0.0--0.1) & 0.6 (0.5--0.6) & 2.1 (0.8--2.0) & & \nodata & \nodata & \nodata \\
48 & 0.9 (0.7--1.0) & 0.6 (0.5--0.6) & 7.7 (3.6--12.4) & & \nodata & \nodata & \nodata \\
50 & \nodata & \nodata & \nodata & & 0.1 (0.0--0.3) & 1.4 (1.0--1.6) & 1.3 (0.0--1.5) \\
53 & \nodata & \nodata & \nodata & & 0.4 (0.1--0.9) & 1.1 (0.6--1.6) & 1.4 (0.0--1.6) \\
54 & \nodata & \nodata & \nodata & & 0.5 (0.3--1.0) & 1.6 (1.2--2.1) & 1.7 (0.0--2.1) \\
64 & 0.8 (0.6--1.0) & 5.7 (3.8--9.9) & 7.1 (3.8--7.4) & & 0.9 (0.7--1.2) & 1.9 (1.7--2.1) & 7.5 (3.6--8.8) \\
66 & 0.4 (0.2--0.7) & 3.2 (2.0--6.4) & 1.3 (0.3--1.7) & & 0.5 (0.3--0.9) & 2.3 (1.8--2.9) & 1.5 (0.0--1.9) \\
68 & 2.1 (1.4--3.1) & 2.0 (1.2--4.7) & 3.9 (0.2--4.3) & & \nodata & \nodata & \nodata \\
71 & 0.8 (0.3--1.4) & 5.7 (2.3--15.0) & 3.3 (0.0--3.7) & & 0.8 (0.5--1.6) & 1.8 (1.3--2.7) & 3.3 (0.0--4.3) \\
72 & \nodata & \nodata & \nodata & & 1.0 (0.8--1.4) & 1.6 (1.3--1.9) & 5.1 (0.0--5.9) \\
  \enddata
 \tablenotetext{a}{The 1 $\sigma$ confidence ranges are given in the parentheses.}
 \tablenotetext{b}{Absorption-corrected values in the 0.5--8.0~keV band. A distance of 300~pc is assumed to derive \lx\ for the thermal fits.}
\end{deluxetable}

\subsection{NIR Data}
All FLAMINGOS frames were reduced following the standard procedures using IRAF; i.e.,
subtraction of dark current, flat-fielding using a dome flat, subtraction of sky using
median sky, removal of bad pixels, and trimming of edges. Several frames with readout
failure were discarded. Artificial signals were masked out, which we found at
pixels to the right and left of, or above and below bright sources by $\sim$128
pixels. Nebulosity extending larger than the dithering amplitude may remain in the
median-sky-subtracted images; it does not affect our photometry. Referring to sources
in the 2MASS all-sky survey catalog \citep{skrutskie97}, all FLAMINGOS frames were
corrected for astrometry by a low-order polynomial function to compensate for the image
distortion. They were also corrected for photometry to match 2MASS magnitudes. After
these corrections, the frames were combined into the final \ji-, \hi-, \ksi-band images.

We extracted sources from the \ksi-band image (Fig.~\ref{fg:f1}\textit{b}) above a
5~$\sigma$ level using SExtractor \citep{bertin96}. The number of false positive
detections is statistically negligible. After manually removing suspicious sources
(introduced mostly by misidentifying outflow knots as point sources) and sources at the
edges, we obtained 294 detections. We measured \ji-, \hi-, and \ksi-band magnitudes at
their positions (Table~\ref{tb:t3}). We matched the FLAMINGOS and the 2MASS sources and
found that all the 67 2MASS sources in the FLAMINGOS field of view have FLAMINGOS
counterparts.

\begin{deluxetable}{lccrrrc}
 \tabletypesize{\scriptsize}
 \tablecaption{FLAMINGOS Photometry\label{tb:t3}}
 \tablecolumns{7}
 \tablewidth{0pt}
 \tablehead{
 \colhead{NIR\tablenotemark{a}} &
 \colhead{R.A.\tablenotemark{b}} &
 \colhead{Decl.\tablenotemark{b}} &
 \colhead{\ji\tablenotemark{c}} &
 \colhead{\hi\tablenotemark{c}} &
 \colhead{\ksi} &
 \colhead{2MASS} \\
 \colhead{ID} &
 \colhead{(hh:mm:ss.s)} &
 \colhead{(dd:mm:ss)} &
 \colhead{(mag)} &
 \colhead{(mag)} &
 \colhead{(mag)} &
 \colhead{Counterpart} 
 }
 \startdata
1 & 3:25:09.4 & 30:48:59 & 18.8 & 18.1 & 17.1 & \nodata \\
2 & 3:25:09.5 & 30:46:21 & 12.5 & 11.0 & 10.1 & 03250943$+$3046215 \\
3 & 3:25:09.7 & 30:48:48 & 13.3 & 12.5 & 12.2 & 03250965$+$3048488 \\
4 & 3:25:10.1 & 30:46:07 & 16.4 & 16.2 & 17.6 & \nodata \\
5 & 3:25:10.2 & 30:47:13 & \nodata & \nodata & 16.5 & \nodata \\
  \enddata
 \tablenotetext{a}{Sources with NIR excess are marked with a dagger.}
 \tablenotetext{b}{Equinox in J2000.0.}
 \tablenotetext{c}{Magnitudes for sources detected with more than 3 $\sigma$ are given.}
 \tablecomments{The complete version of this table is in the electronic
 edition of the Journal.  The printed edition contains only a sample.}
\end{deluxetable}

\begin{figure}[hbtp]
 \figurenum{2}
 \epsscale{1.0}
 \plotone{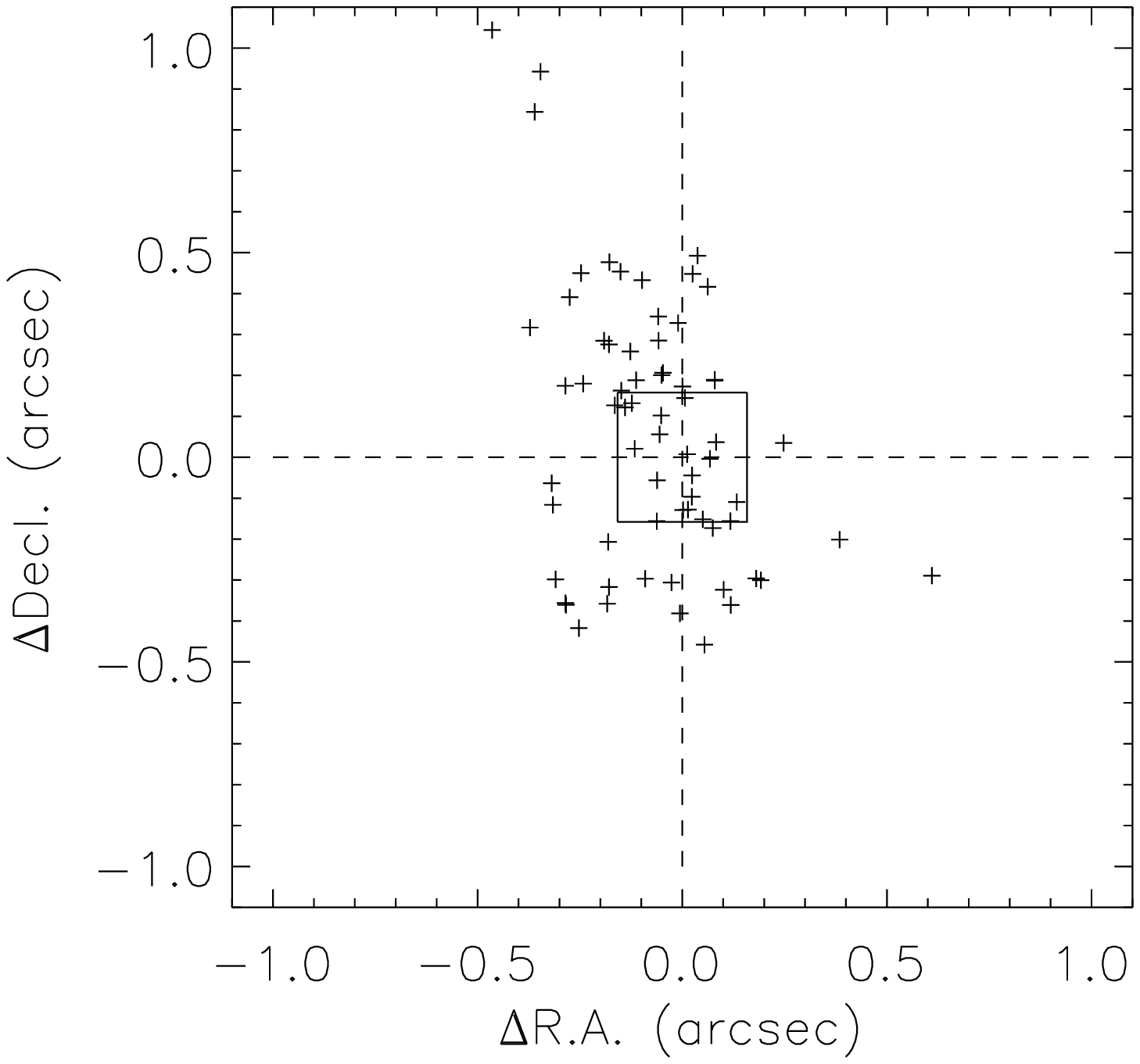}
 \caption{Astrometric accuracy of FLAMINGOS sources. The difference of R.\,A. and
 Decl. between the FLAMINGOS and 2MASS positions ($\Delta$R.\,A. and $\Delta$Decl.) are
 plotted for all the FLAMINGOS--2MASS pairs. The solid square at the center represents
 the FLAMINGOS pixel size of 0.316\arcsec$\times$0.316\arcsec.}\label{fg:f2}
\end{figure}

Figure~\ref{fg:f2} shows the astrometric accuracy of the FLAMINGOS sources, where the
difference of FLAMINGOS and 2MASS positions ($\Delta$R.\,A. and $\Delta$Decl.) are
plotted for the 67 FLAMINGOS--2MASS counterpart pairs. The root mean square of
$\Delta$R.\,A. and $\Delta$Decl. are 0.18\arcsec\ and 0.33\arcsec, indicating that the
FLAMINGOS position accuracy is $\sim$0.4\arcsec\ ($\approx$1.2 FLAMINGOS pixel). The
accuracy is better for most sources, as the values are exaggerated by three sources at
the top left in Figure~\ref{fg:f2}, all of which are located too far away from the image
center to be corrected for distortion.

We estimated the detection completeness of the FLAMINGOS images by embedding artificial
sources and detecting them with the same source detection algorithm. The detection rates
at different magnitude bins were derived to estimate the 90\% completeness limit as
$\sim$18.0~mag, $\sim$16.5~mag, and $\sim$17.0~mag in the \ji, \hi, and \ksi\ bands.

\subsection{NIR and Optical Identification of X-ray Sources}
Seven X-ray sources were found to have FLAMINGOS counterparts within 1.5\arcsec\ and
five additional sources to have 2MASS counterparts outside of the FLAMINGOS view. X-ray
positions in Table~\ref{tb:t1} are shifted by $\sim$0.3\arcsec\ to match the 2MASS
astrometry. The root mean square of the positional difference between 2MASS--X-ray pairs
is $\sim$0.6\arcsec\ ($\approx$1.3 ACIS pixel). Ten of the NIR-identified X-ray sources
are optically identified with the USNO-B catalog \citep{monet03}. None of the
NIR-unidentified X-ray sources were identified in the optical catalog. No X-ray emission
was found from Herbig-Haro objects \citep{bally97} or H$_{2}$ knots \citep{davis95}.

\section{DISCUSSION}
\subsection{Nature of X-ray Sources}
\begin{figure}[hbtp]
 \figurenum{3}
 \plotone{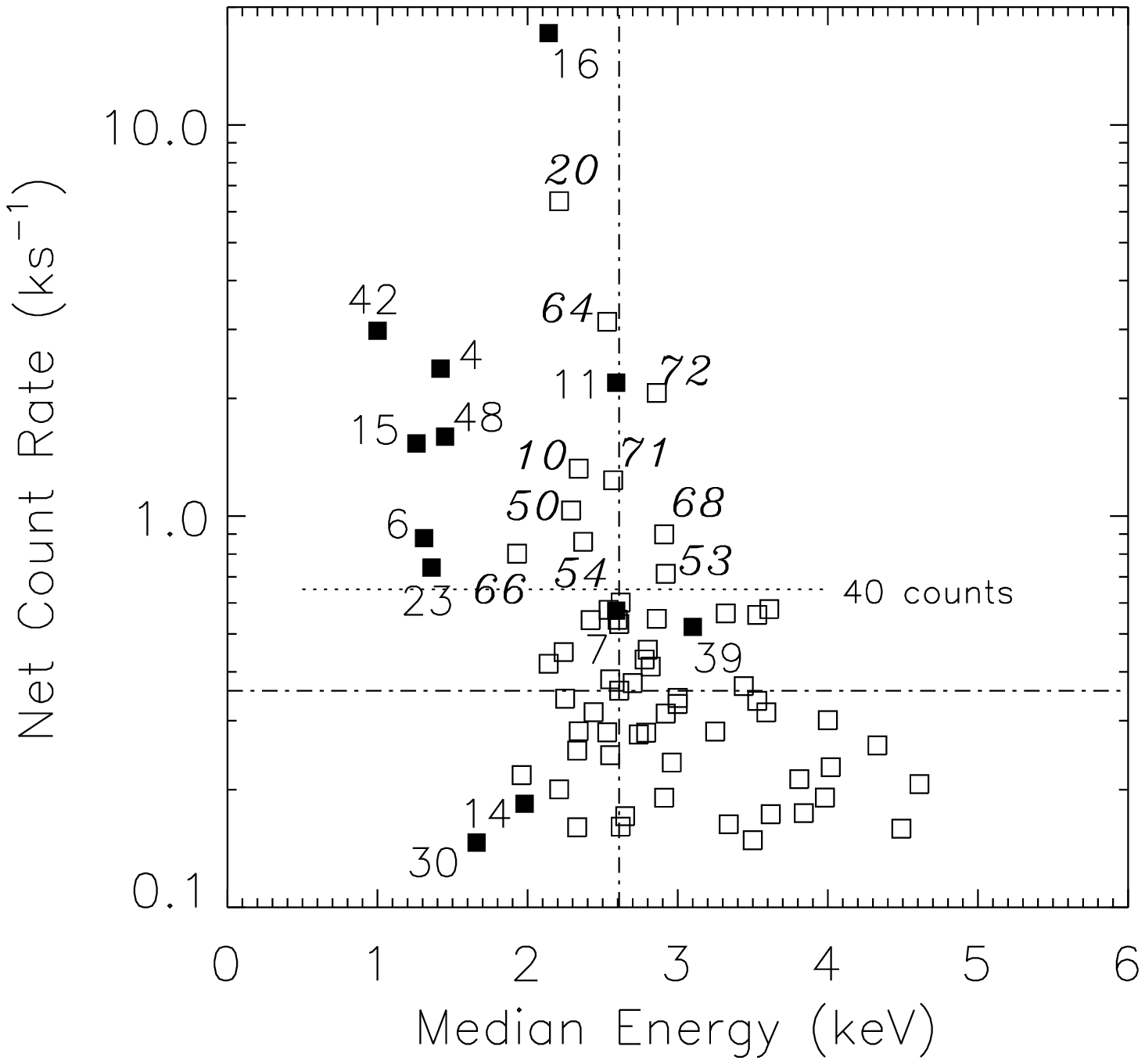}
 \caption{Scatter plot of ME and NCR for all X-ray sources. Sources with and without NIR
 counterparts are shown with filled and open symbols, respectively. The source numbers
 (Table~\ref{tb:t1}) of the former are labeled in Roman, while the latter is in Italic
 if they have $\ge$40 counts. The median value of ME and NCR are shown with
 dashed-and-dotted lines.}\label{fg:f3}
\end{figure}

Two independent quantities from the X-ray photometry (ME and NCR) combined with the NIR
identification give a clue to classify and reveal the nature of the 72 X-ray
sources. Figure~\ref{fg:f3} shows a scatter plot of ME and NCR. Most of the X-ray
sources with NIR counterparts (\textit{filled symbols}) are distributed in the limited
ME range of $\lesssim$2.5~keV, while those without NIR counterparts (\textit{open
symbols}) are distributed in the limited ME and NCR ranges of $\gtrsim$2.0~keV and
$\lesssim$1.0~ks$^{-1}$. This implies that the X-ray sources consist of two major classes
of different nature.

\begin{figure}[hbtp]
 \figurenum{4}
 \plotone{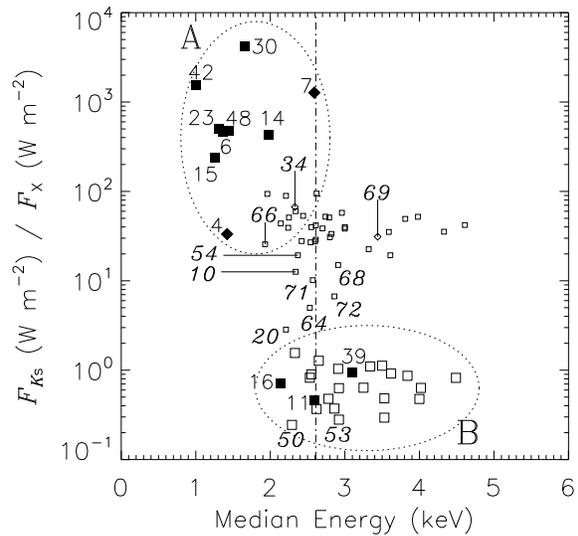}
 \caption{Scatter plot of ME and the flux ratio of NIR (\ksi) and X-ray (0.5--8.0~keV)
 bands for all X-ray sources with and without NIR counterparts (\textit{filled} and
 \textit{open}). Open symbols have either a normal size (for sources inside the
 FLAMINGOS field of view) or a smaller size (otherwise). Variable sources are marked
 with diamonds. Open symbols indicate the upper limit arising from no \ksi-band
 detection. Labels are given in the same manner with Fig.~\ref{fg:f3}. Variable sources
 are also labeled. The median values of ME is shown with the dashed-and-dotted
 line.}\label{fg:f4}
\end{figure}

The separation of the two classes becomes more apparent when NIR flux is taken into
account. Figure~\ref{fg:f4} shows a scatter plot of ME and the ratio of NIR flux in the
\ksi\ band to X-ray flux in 0.5--8.0~keV band. Here, the X-ray flux (\fx) is calculated by
\begin{equation}
 \fx = \frac{\int dE~E~C_{\mathrm{net}}(E)}{\mathrm{EA}~t_{\mathrm{exp}}} = \frac{\mathrm{ME} \times \mathrm{NCR}}{\mathrm{EA}},
\end{equation}
where $t_{\rm{exp}}$ is the exposure time and $\rm{EA}$ is the effective area of the
mirror averaged over the energy. The X-ray sources with NIR counterparts (\textit{large
filled symbols}) and those without FLAMINGOS counterparts (\textit{large open squares})
are clearly separated into two groups. One group (group A) is comprised of X-ray source
Nos.\,4, 6, 7, 14, 15, 23, 30, 42, and 48 in the upper left quarter, while the other
(group B) includes Nos.\,11, 16, 39, and FLAMINGOS-unidentified X-ray sources
(\textit{large open squares}) at the bottom. NIR-unidentified X-ray sources outside of
the FLAMINGOS field of view (\textit{small open symbols}) are hard to classify into
either group as the \ksi-band upper limit of 2MASS is not deep enough. This illustrates
the importance of our NIR observation deeper than 2MASS, even if it results in no
detection from X-ray sources.

\smallskip

We conclude that the sources in group A are mostly YSOs that belong to L1448, while
those in group B are background AGNs. We note that NIR identification of an X-ray source
in star-forming regions does not always guarantee its stellar origin because AGNs, which
constitutes the largest contaminant, can be detected even at the 2MASS depth. Instead,
we employ the NIR to X-ray flux ratio as an effective measure to separate these two
classes. The optical to soft X-ray flux ratio was used for a similar classification
purpose in X-ray blank-sky surveys by \textit{Einstein} and \textit{ROSAT} satellites
\citep{maccacaro88,krautter99}. We use the NIR to total band X-ray flux to be suitable
for star-forming regions, which are routinely highly obscured. In (low-mass) main
sequence and pre--main-sequence (PMS) sources, X-ray emission originates from stellar
coronae while NIR emission originates from photosphere and from circumstellar matter. In
AGNs, on the other hand, X-ray emission is from accretion disks around the central black
hole while NIR is from hot dust and jets. The different origins bring different typical
values of NIR to X-ray flux ratio in stars and AGNs.

Several lines of evidence support this conclusion. First, all sources with $\ge$40
counts in group A (Nos.\,4, 6, 15, 23, 42, and 48) have a successful fit with thermal
spectral models (Table~\ref{tb:t2}) but not with power-law models with reasonable
parameters. On the other hand, all sources with $\ge$40 counts in group B (Nos.\,11, 16,
50, and 53) have a successful fit with power-law models. Three (Nos.\,11, 50, and 53) of
them are not fit with thermal models (Table~\ref{tb:t2}). The clear distinction in X-ray
spectral features is consistent with the idea that (1) these groups of sources are
different in nature, (2) group A sources are stellar X-ray emission, which typically
show thermal spectra of \kt $\sim$ 1~keV, and (3) group B sources are AGNs, which
typically show power-law spectra of $\Gamma \sim$1.7. This conversely enables us to use
X-ray spectral features to classify sources for which it is unclear which group to
belong to. Nos.\,54 and 72 are group B sources for successful fits only by the power-law
model (Table~\ref{tb:t2}).

The second argument is the X-ray flux variability of a typical time scale of
$\sim$10~ks. Stellar X-ray emission from coronae is often characterized by X-ray flares
due to magnetic reconnection, whereas no such short time-scale variability is expected
from AGNs. Sources with X-ray variability are seen in group A (Nos.\,4 and 7) and not in
group B. This conversely indicates that sources with similar variability (Nos.\,34 and
69) belong to group A.

\begin{figure}[hbtp]
 \figurenum{5}
 \plotone{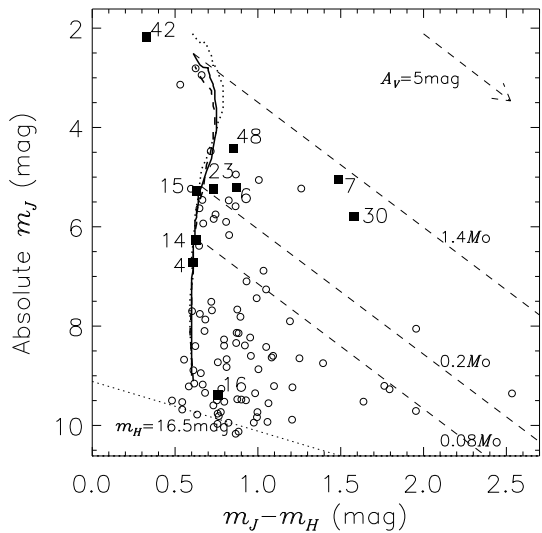}
 \caption{Color-magnitude diagram of X-ray sources with NIR counterparts (\textit{filled
 squares}) and FLAMINGOS sources (\textit{open circles}). Sources with \ji-, \hi-, and
 \ksi-band detections with magnitude uncertainty of $<$~0.05~mag are plotted. Labels are
 given to the X-ray sources with NIR counterparts. The 1, 2, and 3 Myr isochrone curves
 (\textit{dotted, solid, and dashed curves}) are from \citet{baraffe98}. The reddening
 lines for 1.4~$M_{\odot}$, 0.2~$M_{\odot}$, and 0.08~$M_{\odot}$ are given with dashed
 lines assuming an age of 2~Myr.}\label{fg:f5}
\end{figure}

Third, NIR magnitudes of the X-ray sources in group A are distributed in $m_{K_{S}}$ $\sim$
9--13~mag. Contamination by extragalactic sources in this magnitude range is negligible as
the expected number of galaxies in the FLAMINGOS field of view is $\lesssim$1 in $m_{K_{S}}$
$\lesssim$ 14~mag \citep{tokunaga00}. The color-magnitude diagram (Fig.~\ref{fg:f5})
shows that group A sources have consistent NIR magnitudes for low mass or very low mass
stars at a distance of 300~pc. Sources in group B, on the contrary, have $m_{K_{S}}$
$\gtrsim$ 15~mag (Tables~\ref{tb:t1} and \ref{tb:t3}), which is too faint to be stars
at that distance.

Fourth, the NIR to X-ray flux ratio of group A sources (\fk/\fx\ $\sim$
10$^{1}$--10$^{4}$) is consistent with typical values of the X-ray to the bolometric
luminosity ratio of \lx/\lbol\ $\sim$ 10$^{-5}$--10$^{-2}$ \citep{preibisch05} of PMS
sources. Assuming an age of 2~Myr (the same with IC~348; \citealt{muench03}), and using
\lk/\lbol\ $\sim$ 4--9$\times$10$^{-2}$ (0.08--1.4 $M_{\odot}$ PMS source;
\citealt{baraffe98,dantona94}) and \lk/\lx\ $\sim$ \fk/\fx\ $\sim$ 5$\times$10$^{2}$
(the median value of group A sources), we obtain the value of \lx/\lbol\ $\sim$
0.8--1.8$\times$10$^{-4}$ for group A sources. A similar \lx/\lbol\ value is obtained if
we assume that these sources are main sequence stars \citep{tokunaga00,drilling00},
which is higher than their typical values (10$^{-7}$--10$^{-5}$; \citealt{preibisch05}).
We thus consider that most of the group A sources are YSOs in the cloud and not main
sequence stars in the line of sight. The distinction between YSO and main sequence
natures should be confirmed with follow-up spectroscopic studies.  Group B sources, on
the other hand, have \fk/\fx\ values of 10$^{-1}$--10$^{1}$, which is consistent with
the values seen in an X-ray-selected sample of AGNs \citep{watanabe04}.

\smallskip

\begin{figure}[hbtp]
 \figurenum{6}
 \plotone{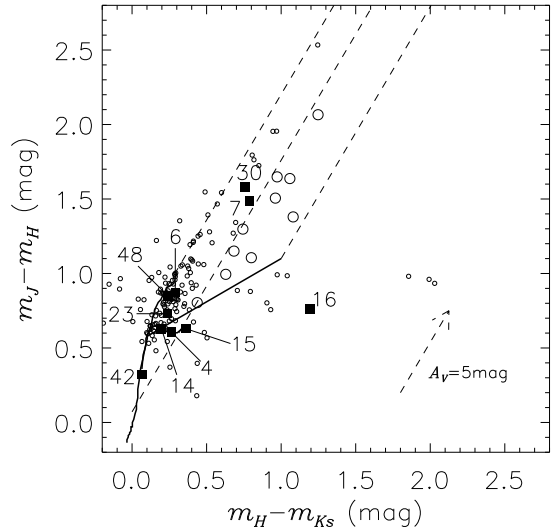}
 \caption{Color-color diagram of X-ray sources with NIR counterparts (\textit{filled
 squares}) and FLAMINGOS sources (\textit{open circles}). Sources with \ji-, \hi-, and
 \ksi-band detections with magnitude uncertainty of $<$~0.1~mag are plotted. FLAMINGOS
 sources with NIR excess emission are shown with larger symbols. The reddening lines are
 shown (\textit{dashed lines}) from the intrinsic colors of dwarfs and giants
 (\textit{thick solid curves}; \citealt{tokunaga00}) and the classical T Tauri star
 locus (\textit{thick solid lines}; \citealt{meyer97}).}\label{fg:f6}
\end{figure}

Figure~\ref{fg:f6} shows a color-color diagram of NIR-identified X-ray
(\textit{squares}) and FLAMINGOS (\textit{circles}) sources. All NIR-identified X-ray
sources in group A (Nos.\,4, 6, 7, 14, 15, 23, 30, 42, and 48) have NIR colors
consistent with being reddened weak-line T Tauri or main sequence stars. None of them
shows apparent NIR excess emission, which does not exclude the possibility that they are
classical T Tauri stars or protostars, as \ksi-band observations are not completely
sensitive to excess emission from circumstellar disks (e.g., \citealt{haisch01}). The
only source with significant \ji-, \hi-, and \ksi-band detections in group B (No.\,16)
has an inconsistently red color for reddened stars of $m_{J}$--$m_{K_{S}}$ $=$ 2.0~mag,
which is commonly seen among 2MASS red AGNs \citep{cutri01}. The density of such sources
is $\sim$0.5~arcdeg$^{-2}$ at \ksi\ $<$ 14.5~mag \citep{cutri01}, making the probability
$\sim$5\% to have a 2MASS red AGN in an ACIS field.

\smallskip

In total, twelve X-ray sources (Nos.\,4, 6, 7, 14, 15, 23, 30, 34, 42, 48, 68, and 69)
have X-ray and NIR features consistent of being low-mass YSOs in this cloud. Two of them
are known sources; No.\,7 is IRS~1 and No.\,42 is an A5 type star \citep{roeser88} that
consists of a double \citep{worley97}. Main sequence A type stars are not expected to
have X-ray emission due to the lack of any plasma creation mechanisms
\citep{schmitt93}. The X-ray emission in No.\,42 may come from a low-mass companion,
which mimics a \fk/\fx\ value consistent with a PMS source (Fig.~\ref{fg:f5}). The
remaining ten sources are candidates of new YSOs in L1448. The brightness histogram of
the rest of the X-ray sources is consistent with a log~$N$--log~$S$ relation of AGNs
\citep{moretti03}.

\subsection{X-ray and NIR Emission from Embedded Sources}
\subsubsection{Infrared Excess Sources}
From Figure~\ref{fg:f6}, ten FLAMINGOS sources (\textit{large open circles}) are found
to have NIR excess emission and thus are candidates for new embedded sources. Daggers are
given for these sources in Table~\ref{tb:t3}. Among them, FLAMINGOS Nos.\,23 and 149 are
promising YSO candidates with estimated magnitude uncertainty of $\lesssim$~0.05~mag.

\subsubsection{IRS~1} 
Both X-ray and NIR emission was found from IRS~1 (ACIS No.\,7 in Table~\ref{tb:t1} and
FLAMINGOS No.\,2 in Table~\ref{tb:t3} or 2MASS 03250943$+$3046215). The X-ray and NIR
properties are consistent with the nature of this source as a classical T Tauri star
or a class~I protostar; (1) it shows X-ray flux variation commonly seen in X-ray
emission of stellar magnetic activity origin, (2) its ME and $F_{K\rm{s}}$/$F_{\rm{X}}$
values indicate this source to be a YSO (Fig.~\ref{fg:f4}), and (3) a nebulous
structure is seen in the FLAMINGOS \ksi-band image (Fig.~\ref{fg:f1}\textit{b}). The
NIR magnitudes of this source ($m_{J}$ $\sim$ 12.5~mag, $m_{H}$ $\sim$ 11.0~mag, and
$m_{K_{S}}$ $\sim$ 10.1~mag) correspond to an object with the mass of
0.5--1.0~$M_{\odot}$ at 300~pc obscured by $A_{\rm{V}} \sim$~7~mag.

\subsubsection{IRS~3(A)}
No emission is seen at the position of IRS~3(A) in our NIR images. The 5$\sigma$ level
of the local sky noise corresponds to $\sim$18.8~mag, $\sim$17.7~mag, and $\sim$17.5~mag
in our \ji-, \hi-, and \ksi-band images. The SED slope defined as $\alpha = (d
\log{\lambda F_{\lambda}}) / (d \log{\lambda}$) between 2.2~\micron\ (\ksi\ band) and
10~\micron\ (\textit{N} band) is commonly used to classify YSOs \citep{wilking89}, using
the fact that the 2--10~\micron\ SEDs at a younger evolutionally stage are more
dominated by emission from circumstellar material. The lower limit of $\alpha$ of
IRS~3(A) is $\gtrsim$3.2 from the non-detection in the NIR bands and a MIR detection in
the \textit{N} band \citep{ciardi03}. Here, we calculated the $\alpha$ value as the
index between \hi\ and \textit{N} bands, as the \ksi-band image is noisy at the position
due to the emission arising from the outflow of IRS~3(A) itself. The value thus is a
lower limit also in the sense that \hi-band emission is subject to contamination from
reflected photospheric emission.

\citet{greene94} gave a qualitative classification of YSOs using $\alpha$, where $\alpha
< -1.6$ for class~III, $-1.6 < \alpha\ < -0.3$ for class~II, and $\alpha > -0.3$ for
class~I sources. Almost all known class~I protostars have $\alpha \lesssim$ 2 with the
only exception of WL~22 ($\alpha =$ 3) in $\rho$ Ophiuchi dark cloud
\citep{wilking89}. Sources with a steeper index than $\sim$ 3 have been rarely known, as
they are quite difficult to detect in \ki\ band images that traditionally pilot
protostar searches. However, recent progress of sensitive MIR imaging both by ground-
and space-based telescopes is changing the situation by reporting several protostars
with very steep spectra with indices of $\alpha>3$ (Cep E by \citealt{noriega-crespo04}
and source X$_{\rm{E}}$ in R CrA by \citealt{hamaguchi05} and private communications with
K. Nedachi). The index of $\gtrsim$3.2 of IRS~3(A) places this source in the class of
very steep spectrum protostars. More sources in this emerging class are expected to be
discovered with \textit{Spitzer} observations of near-by star-forming regions. Some of
them are class~0 protostars (e.g., Cep E mm; \citealt{noriega-crespo04}) and others are
\ki-band-unidentified class~I protostars, or in between.

An unusual concentration of three X-ray photons within a $\sim$0.5\arcsec\ radius
($\sim$90\% encircled energy radius at $\sim$0.4\arcmin\ off-axis angle) can be found at
the position of IRS~3(A). No such concentration is seen within 1\arcmin\ of the on-axis
position. Although it was not recognized as a source in the initial X-ray source search
using the wavelet technique, its PNS value of 5$\times$10$^{-4}$ indicates the existence
of an X-ray source of \fx\ $\approx$ 5$\times$10$^{-16}$~ergs~s$^{-1}$~cm$^{-2}$. The
facts that the position of the concentration is consistent with that of IRS~3(A), that
the energy of all photons is concentrated in the hard (2--5~keV) band, and that the
arrival times of all photons fall in a limited time range of $\sim$15~ks in the
$\sim$68~ks observation, which is indicative of a flare, may further strengthen the
claim. There is no indication of temporal non-uniformity of background counts. A similar
weak but plausible X-ray emission was found in an embedded source (FIR4) in NGC\,2024 by
\textit{Chandra} \citep{skinner03}.

Given the capability of \textit{Chandra} and MIR telescopes (\textit{Spitzer} and
ground-based facilities), it is now becoming possible to investigate X-ray emission from
protostars with a very steep spectrum ($\alpha > 3$) even without \ksi-band
identifications. If IRS~3(A) is indeed an X-ray source, it has the steepest spectral
index among YSOs with X-ray emission. Although it is not yet certain whether a steeper
$\alpha$ value indicates a younger age of protostars, X-ray observations for steep
spectrum sources are a key to answer how early X-ray emission starts and how intense it
is at the earliest stage of star formation. These are important questions to understand
star formation because the coupling between gas and magnetic fields can be largely
affected by X-ray ionization \citep{feigelson99}.

\subsubsection{Class~0 Sources} 
No significant X-ray or NIR emission was found from the four known class~0 sources;
L1448 mm, IRS~2, IRS~3(B), and IRS~3(C). The upper limit (PNS $= 1\times10^{-3}$) of the
X-ray flux in the 0.5--8.0~keV is $\sim$0.5--1.0$\times$10$^{-15}$~ergs~s~cm$^{-2}$,
whereas the upper limits (5~$\sigma$) in the NIR regime are $\sim$18.8~mag (\ji),
$\sim$17.7~mag (\hi), and $\sim$17.5~mag (\ksi), except for L1448 mm and IRS~3(B) in the
\ksi-band, which are contaminated by local diffuse emission.

The lack of X-ray detection from class~0 sources is widely reported. \citet{froebrich05}
compiled 28 well documented bona-fide class~0 sources. Sixteen of them were observed by
\textit{Chandra} and nine in references are reported to have no X-ray detection (L1448
NW and C; this work, NGC\,1333 I2, I4A, and I4B; \citealt{getman02}, OMC\,3 MMS6;
\citealt{tsujimoto02}, HH\,24 MMS and HH\,25 MMS; \citealt{simon04}, and VLA\,1623;
\citealt{imanishi03}). An additional six sources are reported with no detection in
\textit{XMM-Newton} references (IRAS\,03256$+$3055; \citealt{preibisch02}, HH\,211 MM;
\citealt{preibisch04b}, Serpens S68N, SMM2, SMM3, and SMM4; \citealt{preibisch04a}).
Considering that these sources are deeply embedded, the lack of X-ray detection does not
necessarily indicate that they do not have any X-ray-emitting activities, although they
are still difficult for X-ray as well as MIR investigations with current technologies.

\acknowledgments
The authors express gratitude for Kentaro Motohara for his help in the observation at
Kitt Peak, Pat Broos for his advice in X-ray data reduction using ACIS Extract and
proofreading, and Eric D. Feigelson for discussion. We acknowledge financial supported
by the Japan Society for the Promotion of Science (M. T.), a Grant-in-Aid for Scientific
Research of the Ministry of Education, Culture, Sports, Science and Technology
(No. 15740120; Y. T.), a Chuo University Grant for Special Research (Y. T.), the
Saneyoshi Scholarship Foundation (Y. T.), and the \textit{Chandra} guest observer grant
(SAO G04--5009X). FLAMINGOS was designed and constructed by the IR instrumentation group
(PI: R. Elston) at the University of Florida, Department of Astronomy, with support from
NSF grant AST97-31180 and Kitt Peak National Observatory. IRAF is distributed by the
National Optical Astronomy Observatories, which are operated by the Association of
Universities for Research in Astronomy, Inc., under cooperative agreement with the
National Science Foundation. This publication makes use of data products from the Two
Micron All Sky Survey, which is a joint project of the University of Massachusetts and
the Infrared Processing and Analysis Center/California Institute of Technology, funded
by the National Aeronautics and Space Administration and the National Science
Foundation.

Facilities: \facility{CXO(ACIS)}, \facility{KPNO:4m(FLAMINGOS)}

\end{document}